\documentclass[pra,english,reprint, aps,superscriptaddress, nofootinbib]{revtex4-1}
\usepackage[normalem]{ulem}
\usepackage[T1]{fontenc}
\usepackage[latin9]{inputenc}
\usepackage{amsmath}
\usepackage{amssymb}
\usepackage{xcolor}
\usepackage{graphicx}
\usepackage{mathrsfs}
\usepackage{bm} 
\graphicspath{{figures/}}

\definecolor{crimson}{RGB}{164, 16, 52}
\usepackage[pdfusetitle,bookmarks=true,colorlinks=true,
   linkcolor=crimson,
   citecolor=crimson,
   urlcolor =crimson]{hyperref}

\makeatletter

\pdfpageheight\paperheight
\pdfpagewidth\paperwidth

\newcommand{\betavec}{\bm{\beta}}
\newcommand{\alphavec}{{\bm{\alpha}}}
\newcommand{\li}{j} 
\newcommand{\tr}{\text{Tr}}
\usepackage{braket}
\usepackage{comment}

\makeatother

\usepackage{babel}
\begin{document}

\title{Uncovering Universal Entanglement Properties of Correlated Quantum Matter \\in Quantum Simulation} 
  \title{Exploring the Entanglement Structure of Many-Body Wavefunctions \\ on a Quantum Simulator}
\title{Exploring Large-Scale Entanglement in Quantum Simulation} 

\newcommand{\IQOQI}{\affiliation{Institute for Quantum Optics and Quantum Information, Austrian Academy of Sciences, Technikerstra{\ss}e 21a, 6020 Innsbruck, Austria}}
\newcommand{\UIBK}{\affiliation{Institut f\"ur Experimentalphysik, Universit\"at Innsbruck, Technikerstra{\ss}e 25, 6020 Innsbruck, Austria}}
\newcommand{\AQT}{\affiliation{AQT, Technikerstra{\ss}e 17, 6020 Innsbruck, Austria}}
\newcommand{\ITP}{\affiliation{Institute for Theoretical Physics, Technikerstra{\ss}e 21a, 6020 Innsbruck, Austria}}

\author{Manoj K. Joshi}
\email{The first three coauthors contributed equally.}
\IQOQI
\UIBK

\author{Christian Kokail}
\email{The first three coauthors contributed equally.}
\IQOQI
\ITP

\author{Rick van Bijnen}
\email{The first three coauthors contributed equally.}
\IQOQI
\ITP

\author{Florian Kranzl}
\IQOQI
\UIBK

\author{Torsten V. Zache}
\IQOQI
\ITP 

\author{Rainer Blatt}
\IQOQI
\UIBK

\author{Christian F. Roos}
\IQOQI
\UIBK

\author{Peter Zoller}
\IQOQI
\ITP 

\begin{abstract}
Entanglement is a distinguishing feature of quantum many-body systems, and uncovering the entanglement structure for large particle numbers in quantum simulation experiments is a fundamental challenge in quantum information science. Here we perform experimental investigations of entanglement based on the entanglement Hamiltonian, as an effective description of the reduced density operator for large subsystems.  We prepare ground and excited states of a 1D XXZ Heisenberg chain on a 51-ion programmable quantum simulator and perform sample-efficient `learning' of the entanglement Hamiltonian for subsystems of up to 20 lattice sites. Our experiments provide compelling evidence for a local structure of the entanglement Hamiltonian. This observation marks the first instance of confirming the fundamental predictions of quantum field theory by Bisognano and Wichmann, adapted to lattice models that represent correlated quantum matter. The reduced state takes the form of a Gibbs ensemble, with a spatially-varying temperature profile as a signature of entanglement.  Our results also show the transition from area to volume-law scaling of Von Neumann entanglement entropies  from ground to excited states.  As we venture towards achieving quantum advantage, we anticipate that our findings and methods have wide-ranging applicability to revealing and understanding entanglement in many-body problems with local interactions including higher spatial dimensions.
\end{abstract}

\maketitle

\section{Introduction}

\begin{figure*}[t]
	\centering
	\includegraphics[width=0.95\textwidth]{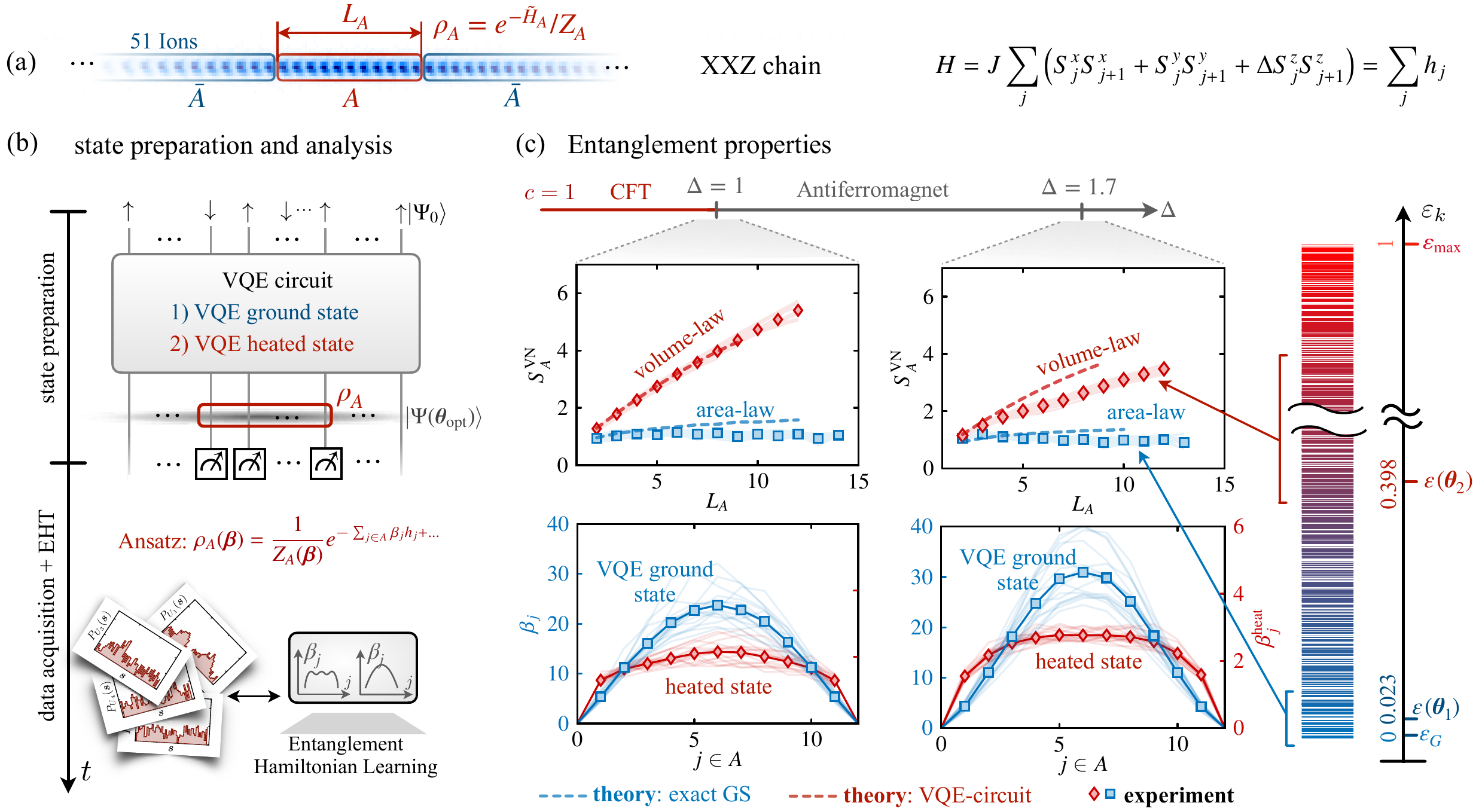}
	\caption{
 \textit{Learning the entanglement structure of variationally prepared ground and heated quantum many-body states.} 
	(a) We study the XXZ model with Hamiltonian $\hat{H}$ given in Eq.~\eqref{eq:xxz}. This defines our EH ansatz $\tilde{H}_A(\boldsymbol{\beta})$ according to Eq.~\eqref{eq:deformation} with $\boldsymbol{\beta}$ the entanglement temperature profile on a subsystem of length $L_A$ in the 51-ion chain. 
 (b) Experimental procedure for state preparation and data analysis. A variational quantum circuit (see Appendix \ref{sec:vqs}) first prepares correlated quantum many-body states. In the second step, we collect frequencies of bit strings sampled in different Pauli bases (see Appendix \ref{sec:eht}). The data are subjected to an entanglement Hamiltonian tomography (EHT) procedure which finds the optimal EH $\tilde{H}_A(\boldsymbol{\beta})$ best reproducing the experimentally obtained frequencies (see Appendix \ref{sec:eht}). (c) Entanglement properties of variationally prepared many-body states on 51 ions, obtained via EHT for different anisotropies $\Delta = 1$ and $\Delta = 1.7$, respectively. Blue squares show the results for the VQE ground states, while red diamonds show data for heated states. The energy spectrum on the left indicates VQE ground and excited state energies as fractions of the entire spectral range. The upper panels show the von Neumann entropies as a function of subsystem size $L_A$ which we obtain from the learned EH via $S_A^\text{VN} = \braket{\tilde{H}_A(\boldsymbol{\beta})} + \log[ Z_A(\boldsymbol{\beta}) ]$. Dashed lines show the corresponding theoretical curves obtained from fitting the ansatz $\rho_A(\boldsymbol{\beta})$ to an MPS simulation of the experiment. Heated-up states indicate a clear volume-law scaling of the entanglement entropy $S_A^\text{VN} \sim L_A$ as opposed to area-law scaling for the VQE ground states $S_A^\text{VN} \sim \text{const}$. Lower panels depict the optimal EH parameters for $L_A = 12$ determined from EHT (see Appendix\ref{sec:eht}). Transparent lines show the results of all connected 12-site subsystems from the central 27 ions of the 51-ion chain. Solid marked lines represent the mean over all these subsystems.  
	}
	\label{fig:EHLearning}
\end{figure*}

Entanglement is the crucial ingredient that sets apart the quantum world from its classical counterpart. 
It is a fundamental concept that has garnered intense research interest due to its implications for various aspects of quantum physics, from foundational aspects to quantum computation to condensed matter systems and quantum chemistry~\cite{doi:10.1142/13369}. In quantum many-body problems, entanglement leads to an exponential scaling of complexity with system size. While classical simulations struggle to capture this complexity, quantum simulation experiments have the ability to naturally represent large-scale entanglement - being quantum systems themselves. Recent years have seen tremendous progress in large-scale quantum simulation experiments touching upon the boundaries of what is classically simulatable~\cite{joshi2022observing,zhangnature2017,justinscience2016,chen2023continuous,semeghini2021probing,sompet2022realizing,leonard2022realization,zhang2022functional}.
The quantum many-body systems in such experiments are typically only locally interacting. That is, operators appearing in the system Hamiltonian act only on local clusters of adjacent particles, with important consequences for the eigenstates of the system and leading to universal features of the entanglement structure contained in them. 

Investigations of bipartite entanglement start with considering a partition of the system of interest into a subsystem $A$ and its complement $\bar A$ (see Fig. \ref{fig:EHLearning} (a)). For a system prepared in a many-body state $\ket{\Psi}$, entanglement between the two can be quantified via the von Neumann entanglement entropy~(EE) $S_A^\text{VN}= - \tr (\rho_A \log  \rho_A)$ where $ \rho_A = \tr_{\bar{A}} \ket{\Psi}\bra{\Psi}$ describes the reduced density matrix of subsystem A. 
For many-body ground states of locally interacting systems, one expects a sub-extensive \emph{area-law} scaling, where the EE only grows with the size of the boundary $\partial A$ of the subsystem~\cite{cirac2021matrix}. This area law scaling lies at the heart of efficient tensor-network approximations in classical simulations of
many-body ground states~\cite{cirac2021matrix,ganahl2023density}. 
In contrast, generic excited states with energies well above the ground state will exhibit \emph{volume-law} scaling, a growth of the EE with the subsystem size reminiscent of the extensive behavior of thermodynamic entropy~\cite{PhysRevLett.127.040603}.  
However, extracting such information about entanglement from large-scale experiments remains challenging, predominantly because of the difficulty of performing tomography of $\rho_A$ on large subsystems~\cite{Acharya_2019} for the purpose of evaluating $S_{A}^{{\rm VN}}$. 
Moreover, identifying quantifiers that capture the entanglement pattern, beyond a simple scalar value $S_{A}^{\rm VN}$, and in relation to the geometry and topology of the subsystem has proven to be a difficult task~\cite{amico2008}.

Here we engage these challenges in an experimental setting by considering the Entanglement (or Modular) Hamiltonian (EH) $\tilde H_A$, which describes the reduced density matrix of a subsystem $A$ through $\rho_A \sim   e^{- \tilde H_A}$. For ground states of many-body Hamiltonians with local interactions, this entanglement Hamiltonian is conjectured to have a simple operator structure. According to fundamental predictions in quantum field theory (QFT) by Bisognano and Wichmann~\cite{BW1,BW2}, the entanglement Hamiltonian can be expressed as a spatial deformation of the system Hamiltonian, i.e., it is composed of the same local operators appearing in the translationally invariant system Hamiltonian, but acquiring a spatially dependent prefactor (see also below, Eq. \eqref{eq:deformation}).
Rather than performing full subsystem tomography of $\rho_A$, we instead `learn' a local entanglement Hamiltonian from experimental data. This allows us to study entanglement properties of large subsystems, and experimentally investigate predictions from QFT. Moreover, the entanglement Hamiltonian provides a unique insight into entanglement patterns by offering an interpretation of $\hat{\rho}_A$ as a Gibbs state with a locally varying inverse temperature, or 'entanglement temperature', quantifying how subregions of the subsystem are entangled with the outside world~\cite{dalmonte2018quantum, giudici2018, zhuprl2020}.

In this work, we prepared ground and excited states of the 1D Heisenberg XXZ model with $N=51$ spins in a trapped-ion platform, and extracted entanglement Hamiltonians for subsystem sizes up to $L_A=20$ lattice sites. 
We find the first experimental evidence for an entanglement Hamiltonian in the form of a deformation of the system Hamiltonian, in line with a fundamental prediction by Bisognano and Wichmann (BW)~\cite{BW1,BW2} and its extension to conformal field theories (CFTs)~\cite{casini2011towards,cardy2016entanglement}.
In addition, our results provide us with the Von Neumann entanglement entropy displaying the area-to-volume law transition~\cite{PhysRevLett.127.040603}. We verify the accuracy of the fitted local entanglement Hamiltonians from independent experimental data, i.e. we assign fidelities to the results without having to resort to theoretical simulations.

\section{Locality of the Entanglement Hamiltonian}

Our study primarily focuses on the entanglement Hamiltonian (EH). In the following discussion, we highlight the remarkable observation that the EH exhibits a local structure in the ground states of quantum field theories (QFTs), as demonstrated in  \cite{BW1, BW2, casini2011towards, cardy2016entanglement, dalmonte2022entanglement}. Moreover, this fundamental finding has been extended to lattice models \cite{dalmonte2022entanglement}, providing insights into the operator content of the EH in locally interacting quantum many-body systems. This allows for efficient protocols to comprehend the EH, which serves as the foundation for measuring entanglement in large subsystems within our quantum simulation experiments \cite{EHT}.

Relativistic quantum field theory (RQFT) in $d+1$ dimensional Minkowski
space makes  predictions about the entanglement
structure of the vacuum (ground) state $\ket{\Omega}$. The Bisognano and Wichmann theorem~\cite{BW1,BW2} states that the reduced density operator of the ground state when partitioned into a semi-infinite half-space $A = \{\boldsymbol{x} \in \mathbb{R}^d \;|\; x_1 > 0\}$ and its complement $\bar{A}$, takes the form of a Gibbs state, 
\begin{align}\label{eq:BWQFT}
\rho_{A}\equiv{\rm Tr}_{\bar{A}} \ket{\Omega}\bra{\Omega} =\frac{1}{Z_A}e^{-\int_{A} {\sf d}^{d}x\,\beta(\boldsymbol{x}){\mathscr{H}}(\boldsymbol{x})}\equiv \frac{1}{Z_A}e^{-\tilde{H}_{A}}
\end{align}
Here, ${\mathscr{H}}(\boldsymbol{x})$ is the Hamiltonian density of the
RQFT, and $\beta(\boldsymbol{x})\sim x_{1}$ is the inverse temperature
profile, that follows a linear ramp with $x_1$ being the coordinate perpendicular to the plane $\partial A$ cutting the infinite half-spaces. This result generalizes to conformal field theories (CFTs)~\cite{casini2011towards,cardy2016entanglement}, predicting in particular that for a region $A$ as a ball of radius $R$ and radial coordinate $r = |\boldsymbol{x}|$, the inverse temperature profile has the form of a parabola, $\beta(\boldsymbol{x})\sim (R^{2}-r^{2})/2R$ (see Appendix for details).

Notably, Eq.~\eqref{eq:BWQFT} implies that the entanglement Hamiltonian $\tilde{H}_{A}$ is local and shares the same operator structure as the system Hamiltonian but is spatially deformed according to the profile $\beta(\boldsymbol{x})$ defining an `entanglement temperature' $T(\boldsymbol{x})= 1/\beta(\boldsymbol{x})$. This temperature decreases with increasing distance from the cut, indicating that the dominant contributions to the entanglement, due to low-lying eigenfunctions of $\tilde{H}_A$, are supported close to the cut.

By adapting and applying BW arguments to the EH for ground states of strongly interacting lattice models with local interactions, a conjecture has emerged that generalizes Eq.~(1)~\cite{dalmonte2022entanglement}. 
This conjecture has been supported by both numerical and analytical investigations across various many-body models. To elaborate, let's consider a spatial deformation of a lattice Hamiltonian:
\begin{align}\label{eq:deformation}
\hat H = \sum_{j}\hat{h}_{j} \quad \overset{\text{deform}} {\xrightarrow{\hspace*{23pt}}} \quad \tilde{H}_A = \sum_{j\in A} \beta_j \hat h_j + \dots,
\end{align}
where $\hat{h}_{j}$ represents few-body terms that act on a neighborhood of lattice sites $j$, and $\beta_j$ denotes the deformation profile. According to the conjecture, the ground state of $\hat H$ gives rise to a reduced density operator $\rho_A$ for a connected subsystem $A$, which assumes the form described in Eq.~(\ref{eq:BWQFT}), with $\tilde{H}_A$ derived from Eq.~(\ref{eq:deformation}). It is expected that this statement holds true for states that can be effectively described by a continuum field theory, to which the BW theorem can be applied, with minor non-universal corrections indicated by the dots.

In the current context, these arguments suggest a general operator structure for the entanglement Hamiltonian that can be experimentally explored. By parametrizing $\tilde{H}_A$ as in (\ref{eq:deformation}) and testing for potential deviations, we employ a learning protocol called EHT \cite{EHT,anshu2021sample}, owing to the locality of $\tilde{H}_A$. In other words, the number of samples required to learn the coefficients $\beta_j$ within a given error scales polynomially with the number of terms $\hat{h}_j$.

Simultaneously, this procedure enables the direct measurement of the Von Neumann entanglement entropy (EE), given by $S_A^{\rm VN}= {\rm Tr} ( \rho_A \tilde{H}_A) +\log Z_A$. Furthermore, this methodology can also be applied to more general excited or thermal states. In the latter case, a Gibbs state with a flat inverse temperature profile $\beta=\beta_j$ is expected (with boundary corrections~\cite{GroverPRX}). Extracting the EH for states at various energies, spanning from the lowest to the middle of the spectrum of $\hat H$, facilitates the experimental observation of the transition from an area law to a volume law for the EE.

\section{Experimental setup and model}

A programmable trapped-ion quantum simulator serves as our experimental platform for studying the entanglement structure in correlated quantum many-body states. In our setup, a linear chain of $N=51\  ^{40}$Ca$^{+}$ ions is held in a linear Paul trap using highly anisotropic confining potentials. The spin states are encoded into long-lived electronic states $\ket{\downarrow} = \ket{S_{1/2},m=+1/2}$ and $\ket{\uparrow} = \ket{D_{5/2},m=+5/2}$,  defining the computational basis. Global entangling operations are realized via quench dynamics of an XY model with controllable long-range interactions, which is engineered via exploiting the spin and motional degrees of freedom of the trapped ions (see Appendix \ref{sec:Jijmatrix}). Our system allows spatially-resolved addressing and detection, enabling arbitrary single-qubit rotations and high-fidelity readout (see Appendix \ref{se:StatePrep}).

To study universal features of the entanglement structure on the trapped-ion platform, we focus on realizing low-energy states of the Heisenberg XXZ model with open boundary conditions
\begin{align} \label{eq:xxz}
\hat{H}_\text{XXZ} = J \sum_{j = 1}^{N-1} \left( \hat{S}_{j}^x \hat{S}_{j+1}^x + \hat{S}_{j}^y \hat{S}_{j+1}^y + \Delta \hat{S}_{j}^z \hat{S}_{j+1}^z \right),
\end{align}
where $\hat{S}_{j}^\alpha $ denote spin-1/2 operators acting on lattice sites $j$. For an anisotropy parameter $-1 < \Delta \le 1$, the critical regime, this model is gapless and its low-energy physics is described by a CFT with central charge $c \!=\! 1$~\cite{calabrese2004entanglement}. 
Preparing low-energy states in this regime allows us not only to experimentally study lattice analogs of the BW theorem 
on half partitions, but also finite bulk intervals.
We further present results analyzing the change of the temperature profile outside the critical regime $\Delta > 1$ as well as for highly excited states of the XXZ chain. 

\section{Results}

\begin{figure*}[t]
	\centering
	\includegraphics[width=0.9\textwidth]{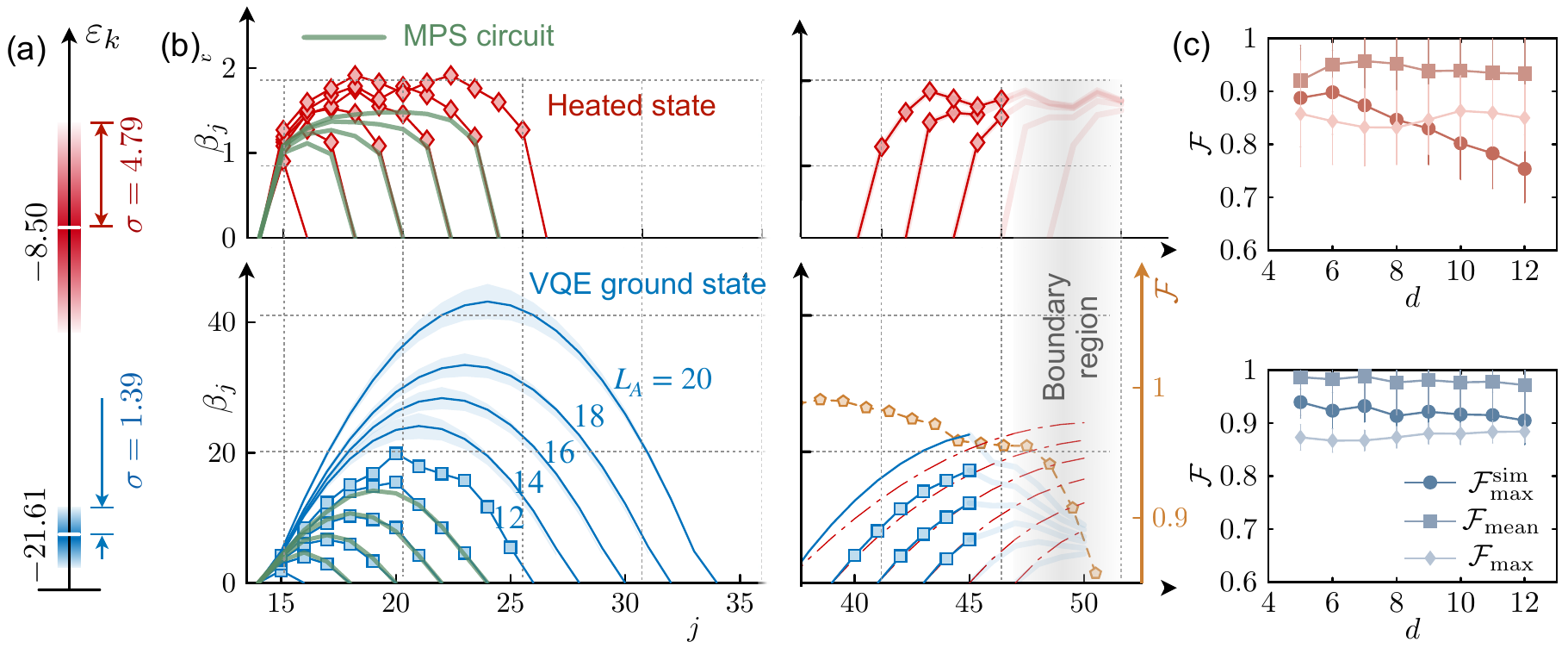}
	\caption{\textit{Entanglement temperature profiles for different subsystem sizes at the critical point $\Delta = 1$} (a) Energy distribution of VQE states calculated from the mean and variance of $\hat{H}$ in the corresponding MPS wave functions. (b) Local inverse temperatures $\beta_j$ for different subsystem sizes in the bulk and at the boundary of the 51-ion chain obtained from EHT (see Appendix). The lower panels (blue curves) show the results for the VQE ground state up to $L_A=20$ sites. Temperature profiles up to $L_A = 12$ sites are obtained from an EH ansatz with local fit parameters $\beta_j$ (blue squares) and operator components $\hat{h}_j$ as defined in Fig.~\ref{fig:EHLearning}~(a). For $L_A > 12$, we describe the temperature profile with a second-order polynomial $\beta_j = q_0 + q_1 j + q_2 j^2$, introducing 3 global fit parameters $\{q_m\}_{m=0}^2$. 
 The orange pentagons plotted in the lower right panel of (b) shows the Uhlmann fidelity of the learned $\rho_A(\boldsymbol{\beta})$ with respect to the corresponding $\rho_A$ from the exact ground state for subsystems of size $L_A = 7$ that we sweep through the ion chain.
 The observed fidelity drop at the edges of the chain causes the learned coefficients $\beta_j$ to deviate from the ones of the exact ground state (red dash-dotted lines) in a boundary region. The coefficients $\beta_j$ observed in the heated state (red diamonds) are consistent with the uniform temperature profile of a thermal Gibbs state with a notable decrease of the local inverse temperature compared to the VQE ground state. (c) Verification of the EHT procedure via cross-fidelity estimation (see Appendix \ref{Methods:XP}). Reduced density matrices of size $L_{A'} = 5$ are computed from the learned $\rho_A(\boldsymbol{\beta})$ and cross-verified against independent data taken from the experiment. Circles represent the maximum fidelity $\mathcal{F}_\text{max}^\text{sim}$ with respect to theoretical simulations. Error bars have been obtained via Jack-knifing and are smaller than symbols if not shown.
	}
	\label{fig:BW}
\end{figure*}

\begin{figure*}[t]
	\centering
	\includegraphics[width=0.95\textwidth]{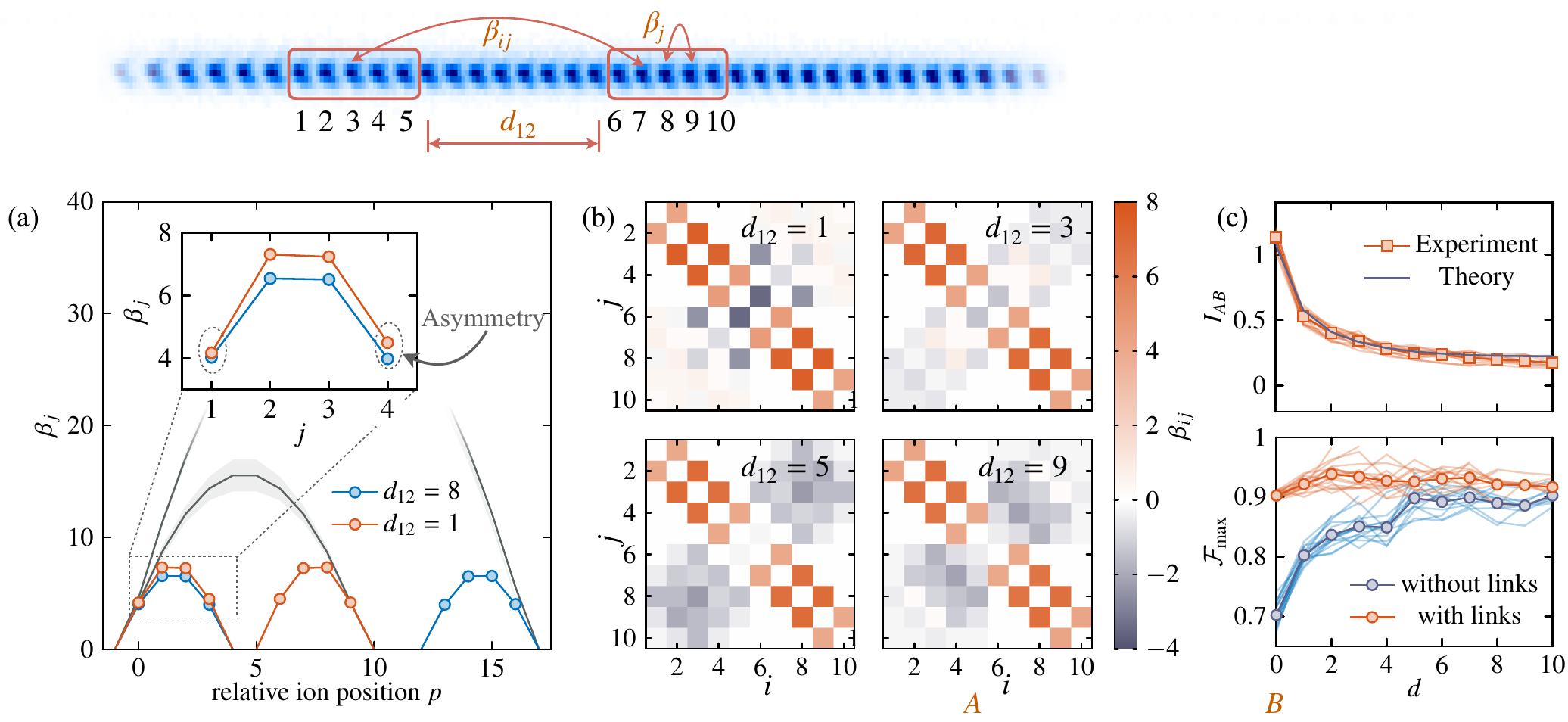}
	\caption{\textit{Entanglement Hamiltonian of disjoint 5-site subsystems.} (a) Local temperature profiles of disconnected subsystems with separations $d_{12} = 1$ (orange) and $d_{12} = 8$ (blue). The inset indicates the increased inverse temperatures $\beta_j$ of subsystems with separations $d_{12} = 1$ as opposed to subsystems with a separation $d_{12} = 8$, resulting in the decrease of mutual information as demonstrated in panel (b). The inset further illustrates an asymmetry of the temperature profiles which we interpret as a temperature gradient (as numerically observed in \cite{sels2017quantum} in the context of free-fermion models). (b) Colormaps of the EH parameters $\beta_{ij}$ for different subsystem separations. (c) Upper panel: mutual information $I_{AB} = S_A^{\text{VN}} + S_B^{\text{VN}} - S_{AB}^{\text{VN}}$ as a function of subsystem separation $d_{12}$ together with theoretical predictions from the MPS wave function. Lower panel: maximum fidelity with respect to an independent data set as a function of subsystem separation $d_{12}$, averaged over many subsystems. The orange lines depict the fidelity of an ansatz $\tilde{H}_A(\boldsymbol{\beta})$ which includes cross-links $\beta_{ij}$ with $i \in A$ and $j \in B$, while the blue line uses the standard ansatz for two independent subsystems. For small separations, cross-links $\beta_{ij}$ lead to a significant boost in fidelity. Error bars are smaller than symbols} 
	\label{fig:disc}
\end{figure*}

Our main experimental procedure consists of variational state preparation followed by Entanglement Hamiltonian Tomography as outlined in Fig.~\ref{fig:EHLearning}~(b). 
We performed experimental preparations of approximate ground and excited states for an XXZ model (cf.~Eq.~(\ref{eq:xxz})) by optimizing quantum circuits generated through quench dynamics in a variational quantum eigensolver (VQE) feedback loop \cite{kokail2019self}. This resulted in variational states that correspond to superpositions of eigenstates from finite energy windows within the XXZ model (indicated by the blue and red brackets in the energy bar of Fig.~\ref{fig:EHLearning}~(c)).

We successfully prepared states with an energy distance from the true ground state that corresponds to approximately 2\% of the entire spectral range (see Appendix \ref{sec:vqs}). Furthermore, states with high average energy, located in the middle of the spectrum, can be prepared efficiently by quenching the initial state using the native interaction Hamiltonian of the ion chain (see Appendix \ref{sec:Jijmatrix}), and subsequently applying the same VQE circuit utilized for preparing low-energy state. We will refer to the states prepared using this method as `VQE heated states'.

Subsequently, we analyze the entanglement properties of the prepared many-body states via EHT, using experimental samples from 243 Pauli bases as base data. In each of these Pauli bases, we collect 200 samples from quantum projective measurements. EHT is performed from an ansatz of the EH of the form $\tilde{H}_A(\boldsymbol{\beta}) = \sum_j \beta_j \hat{h}_j$ with operator components $\hat{h}_j$ defined in Fig.~\ref{fig:EHLearning} (a). The EHT procedure is independently verified via cross-fidelity check with respect to independent experimental data sets and theoretical simulations (see below).

\paragraph{Area-law and volume-law scaling of the entanglement entropy.} Our main results are summarized in Fig.~\ref{fig:EHLearning}~(c), in which we analyze the entanglement structure of VQE ground and heated states for different values of the anisotropy $\Delta$. 
For subsystems $A$ in the bulk of the chain, we observe the anticipated distinct behavior for the low-energy and excited states. While the former exhibits an approximately constant EE, consistent with an area law of entanglement, the heated-up states exhibit a characteristic growth of the EE, which is consistent with a linear, volume law, scaling $S_A \propto L_A$ with subsystem size $L_A = 2, \dots, 12$. This behavior is intimately related to the characteristic shape of the EH temperature profiles $\beta_j$, displayed in the lower panels of Fig.~\ref{fig:EHLearning}~(c). The parabolic shape of the profiles in the VQE ground state results in spins near the boundary of $A$ providing the dominant contribution to the entanglement with the environment $\bar{A}$, thus capturing the essential feature of area-law entanglement. In contrast, the profiles for the excited states $\beta_j$ exhibit a relatively flat plateau within the bulk of the subsystem, with only small differences observed between the boundary and bulk spins. In either case, we find a smooth profile for the learned parameters, consistent with expectations from CFT.

Furthermore, our results provide indications of a distinctive behavior between temperature profiles of the VQE ground states in the critical and non-critical regimes. As can be seen in the lower right panel of Fig.~\ref{fig:EHLearning}~(c) ($\Delta = 1.7$), the flanks of the profile $\beta_j$ exhibit an approximately constant slope, to be contrasted to the parabolic profile in the critical regime ($\Delta = 1$). Our findings are consistent with the analytical results of free-particle systems \cite{Eisler_2020}, where EH parameters of non-critical chains follow a triangular deformation. 
This suggests that the different shapes (parabolic vs.~triangular) of the EH parameters reflect the distinct functional behavior in the decay of correlation functions (power-law vs. exponential).

\paragraph{Scaling behavior of entanglement temperature profiles with $L_A$.} We now turn to analyzing the entanglement structure in more detail, by studying the behavior of the EH as a function of subsystem size for bulk and boundary regions at the critical anisotropy $\Delta = 1$. In Fig.~\ref{fig:BW}~(a), we display normalized variance and mean energies for variationally prepared ground and the excited states for which we summarise the results below. 
Fig.~\ref{fig:BW}~(b) summarizes the resulting entanglement temperature profiles $\beta_j$, encoding the information of subsystem density matrices of the 51-ion chain up to $L_A = 20$ sites. For the VQE ground state, the profiles exhibit a parabolic shape for all subsystem sizes $L_A$ whose height grows approximately linearly with $L_A$, i.e.~individual spins that are close to the subsystem's edges contribute dominantly to the entanglement. 
Although the original BW predictions are made for ground states, we find that even for the approximate ground states, which are superpositions of states spanning a finite energy range of the spectrum, the parabolic profile remains robust. In Appendix \ref{sm:200}, we show numerically for the lowest 200 states that each individual eigenstate exhibits a parabolically deformed EH. 
The heated-up states however, are much higher in energy and here the inverse entanglement temperature profile flattens and forms a plateau, whose height stays approximately constant as a function of subsystem size, resulting in a linear scaling of entanglement entropy (see Fig.~\ref{fig:EHLearning}~(c)) reminiscent of a locally thermalized state.

The BW theorem predicts a linear slope of the temperature profile $\beta_j \sim j$ for a bi-partition of an infinite system into two halves, and we expect this profile to bend over in the presence of a boundary \cite{EHT}.
These expectations are confirmed by our EH learning procedure for the VQE ground state as illustrated for subsystems at the edge, mimicking a half-infinite subsystem, of the ion chain in Fig.~\ref{fig:BW}~(b), in agreement with exact ground state simulations (red dashed-dotted line). 
The profiles close to the boundary of the ion chain, however, reveal that VQE state preparation is less accurate 
there, which we attribute to finite-size artifacts due to the low depth of our variational circuit.
In order to quantify this effect, we compute the average fidelity of 7-qubit reduced density matrices compared to the corresponding subsystems in the exact ground state (see lower right panel of Fig.~\ref{fig:BW} (b)), which shows a clear 
deviation from the exact ground state in the boundary region. We note that the primary contribution to the density matrix stems from spins located in close proximity to the entanglement cut, which clarifies the slight drop in geometric mean fidelity (see Appendix \ref{Methods:XP}) at the boundary.

\paragraph{Verification.}
We perform direct experimental verification of the reconstructed density matrices, by employing protocols similar to those described in Refs.~\cite{ElbenXPlatform,elben2023randomized}. 
The procedure we employ works as follows. Having obtained $\rho_A(\boldsymbol{\beta})$ for a given subsystem $A$, we further split the subsystem $A$ into 2 subintervals $A_1$ and $A_2$, and compute reduced density matrices for the subinterval $A_1$ via $\rho_{A_1}(\boldsymbol{\beta}) = \text{Tr}_{A_2} \left[ \rho_A(\boldsymbol{\beta}) \right].$ We then cross-verify the model $\rho_{A_1}(\boldsymbol{\beta})$ against an independently taken data set via a Hilbert-Schmidt fidelity estimation (see \cite{ElbenXPlatform} and Appendix \ref{Methods:XP} for details) and average the resulting fidelity over all connected subintervals $A_1$ of $A$. Fig.~\ref{fig:BW}~(c), summarizes the results of this verification procedure up to $L_A = 12$ sites, for subinterval sizes $N_{A_1} = 5$. For both VQE ground state and heated state, the geometric mean fidelity $\mathcal{F}_\text{mean}$ is significantly above 90~\%, while the max fidelity  $\mathcal{F}_\text{max}$ in the heated state approaches $\sim$85\% (as defined in Appendix \ref{Methods:XP}). 

Moreover, we perform the same verification procedure not only with respect to independent data sets from the experiment but also to data from theoretical simulations. To this end, we simulate the variational circuits using a time-dependent variational principle with Matrix Product States (MPS) on 51 spins and compute reduced density matrices of subintervals $\rho_{A_1}^\text{MPS}$ from the MPS wave function, and with these, we perform direct fidelity estimation with the experimental data. The results, denoted as $\mathcal{F}_\text{sim}$ in Fig.~\ref{fig:BW}~(c), demonstrate that the maximum fidelities between $\rho_{A_1}(\boldsymbol{\beta})$ and $\rho_{A_1}^\text{MPS}$ are consistent with the experimental fidelities within the error bars.

\paragraph{Entanglement structure of disjoint subsystems.}
 So far, we have focused on a single connected subsystem and demonstrated the universal applicability of the BW arguments.  We now investigate \emph{disconnected} subsystems $A \cup B$ with regions $A$ and $B$ that are separated by a distance $d_{12}$, where no universal predictions are available. 
We find that the reduced density operator is well captured by an EH of the form
\begin{align}\label{eq:disc_ansatz}
\tilde{H}_{A\cup B} = \sum_{ij \in A\cup B} \beta_{ij} \; \hat {\boldsymbol{S}}_i \cdot \hat {\boldsymbol{S}}_j\;,
\end{align}
where $\hat{\boldsymbol{S}} = \left(\hat{S}^x, \hat{S}^y,\sqrt{\Delta} \hat{S}^z\right)^T$ and $\beta_{ij} = \delta_{i,j-1} \beta_i$ whenever $i$ and $j$ are within the same sub-subsystem.

We analyze disconnected subsystems for the $\Delta = 1$ dataset. As shown in Fig.~\ref{fig:disc}~(a), the intra-subsystem profiles $\beta_j$ approach the expected parabolic shape for large separations, indicating that $A$ and $B$ become statistically independent. 
For short distances, these profiles are modified and acquire an asymmetry, in qualitative agreement with analytic predictions for specific CFTs~\cite{casini2009reduced,eisler2022local} and reminiscent of an entropic force~\cite{sels2017quantum}. 
We further quantify this effect by calculating the mutual information $I_{AB}$ shown in Fig.~\ref{fig:disc}~(c), which increases for small $d_{12}$ and approaches a small constant value for large $d_{12}$, in agreement with our theoretical simulations. 
The correctness of the ansatz~\eqref{eq:disc_ansatz} is verified by again computing the fidelity $\mathcal{F}_\text{max}$, which exceeds $90\%$, for all values of $d_{12}$. 
Omitting the additional terms in the ansatz that connect the subsystems A and B leads to markedly lower fidelities (see Fig.~\ref{fig:disc}~(c).
The values of all necessary fit parameters $\beta_{ij}$ are shown in Fig.~\ref{fig:disc}~(b), where the emergence (vanishing) of inter-subsystem coupling with decreasing (increasing) distance becomes apparent.

To summarize, despite the small size of the subsystems studied here, our findings provide the first experimental evidence in favor of bi-local generalization of Eq.~\eqref{eq:deformation}, compatible with a CFT calculation for a massless Dirac field~~\cite{casini2009reduced,eisler2022local}. We again emphasize that a general prediction for the EH for disconnected subsystems is presently not available. Applying our approach to different models with a known effective CFT description can therefore help to improve our understanding of entanglement properties for general CFTs.

\section{Conclusions and Outlook}

The entanglement Hamiltonian provides a powerful tool to study entanglement in correlated quantum matter governed by local Hamiltonians. In case the EH is local, it is not only efficiently learnable from experimental data, but it also provides a readily interpretable 'entanglement temperature' profile providing insights into the entanglement structure of the underlying quantum many-body state.
Our work presents the first experimental observation of a local EH in strongly interacting lattice models, as an extension of predictions from BW originally made in the context of ground states of QFTs. Interestingly, we observe that the local structure of the EH is robust and persists over a large range of low-energy states. We have studied low and high energy states of the Heisenberg model with 51 spins in a trapped ion quantum simulator. The local operator structure of the learned EHs is verified by direct fidelity estimation from independent data, as well as numerical simulations providing excellent agreement. Our methods also enable clear observation of the transition from an area law of entanglement to a volume law in excited states, with entanglement temperature profiles transitioning from strongly deformed to near-uniform distributions.

We anticipate that the entanglement characteristics of ground states explained by BW arguments apply to a broad class of many-body systems with local Hamiltonians including higher spatial dimensions and fermionic systems. Furthermore, the toolset used for measuring the operator structure of the entanglement Hamiltonian can be applied to all present-day programmable quantum simulation platforms. These advancements provide a framework for exploring entanglement-related phenomena in experiments systematically. For instance, the approach can be used to recognize topologically ordered phases of matter through entanglement spectroscopy \cite{zache2022entanglement,li2008entanglement,chandran2014universal}, or it can be used to test new concepts providing further insights into entanglement structure~\cite{murciano2022negativity,kim2022modular}. Furthermore, the Ryu-Takayanagi conjecture \cite{ryu2006holographic} quantitatively relates entanglement properties of CFTs to the geometry of a dual gravitational theory, enabling the indirect study of gravity through the holographic principle, where recent quantum simulation experiments~\cite{periwal2021programmable} provide the necessary programmability of interactions. A shift of focus to the EH as the central object of study in investigations of entanglement in many-body systems thus opens the door to a broad class of new physics to be explored on programmable quantum simulators.

\small

\section*{Acknowledgements} CK and PZ thank Dries Sels for discussions. This project has received funding from the European Union's Horizon 2020 research and innovation programme under grant agreement No 101113690 (PASQuanS2.1). CK, RvB, TVZ, and PZ were supported by
the US Air Force Office of Scientific Research (AFOSR) via IOE Grant No. FA9550-19-1-7044 LASCEM, 
the Austrian Research Promotion Agency (FFG) contract 884471 (ELQO, RvB), and by the Simons Collaboration on Ultra-Quantum Matter, which is a grant from the Simons Foundation (651440, PZ). MJ, FK, CFR, and RB acknowledge the financial support for the experiment from the Austrian Science Fund through the SFB BeyondC (F7110), and the Institut f\"ur Quanteninformation GmbH. This research was supported in part by the National Science Foundation under Grant No. NSF PHY-1748958.
The computational results presented have been achieved (in part) using the HPC infrastructure LEO of the University of Innsbruck. Simulations were performed using  iTensor~\cite{itensor}.

\section*{Author contribution} MJ and FK developed and conducted the experiment under the guidance of RB and CR. CK, RvB, TVZ, and PZ proposed the research and developed the quantum protocols. CK, RvB, and MJ performed the data analysis. CK, RvB, TVZ, and PZ wrote the manuscript, and MJ contributed texts on experimental setups. All authors contributed to the discussion of the results. 

\appendix

\section{Characterizing bi-partite entanglement}
\label{sec:CBE}

The main text considers bipartite entanglement properties of the ground and excited states. For a quantum system in a pure state $\ket{\Psi}$, the 
entanglement properties with respect to a  bipartition \mbox{$A\!:\!\bar A$} are specified by the Schmidt decomposition, $\ket{\Psi}=\sum_{\alpha=1} ^{\chi_A} \lambda_\alpha \ket{\Phi^\alpha_{A} } \otimes \ket{\Phi^\alpha_{\bar A} }$. 
Here, $\lambda_\alpha$ are Schmidt coefficients, and the Schmidt rank $\chi_A$ serves as a proxy of entanglement. 
Defining a reduced density matrix ${\rho}_{A}={\rm Tr}\left[ \ket{\Psi}\bra{\Psi}\right]$, and an EH $\tilde H_A$ as in Eq.~\eqref{eq:BWQFT}, we identify the Schmidt vectors $\ket{\Phi^\alpha_{A} }$ with the eigenvectors of $\rho_A$ and $\tilde H_A$. The entanglement spectrum (ES) is defined via $\lambda_A^2 = e^{-\xi_\alpha}$. Thus, knowledge of $\rho_A$, or equivalently $\tilde H_A$ -- as provided by sample-efficient learning of the entanglement Hamiltonian in Appendix \ref{sec:eht} -- fully specifies bipartite entanglement.

Area and volume law scaling are defined via the bipartite Von Neumann entanglement entropy  defined  as $S^{\rm VN}_A~=~- {\rm Tr} \left( \rho_A \log \rho_A\right)$, or in terms of the EH as $S_A^{\rm VN}~=~{\rm Tr} (\rho_A \tilde H_A )  + \log Z_A$. Area law behavior, as is characteristic for many-body ground states,  is identified as the scaling $S^{\rm VN}_{A}\propto L_{A}^{d-1}$ (or $\log L_{A}$
for $d=1$ at critical points), where $L_{A}$ is the linear extent
of the subsystem $A$ and $d$ denotes the number of spatial dimensions. In contrast, volume law scaling is given by $S_{A}\propto V_A=L_{A}^{d}$, as expected, e.g.~for a thermodynamic entropy. For the $d=1$ Heisenberg model, area and volume law scaling are observed for (approximate) ground and excited states, respectively, in Fig.~\ref{fig:EHLearning}~(c).

\section{Approximated power-law interactions}
\label{sec:Jijmatrix} The experimental platform is discussed in previous references \cite{Kranzl2022,joshi2022observing}. In our trapped-ion quantum simulator, an approximated power-law type spin-spin interaction is engineered by laser fields driving the electronic and transverse motional degrees of freedom of the trapped ions, allowing us to perform entangling steps in the variational optimization loop. The interaction is described by 
\begin{align} \label{eq:xy}
\hat{H}_{\text{XY}} = \sum_{i,j>i}  J_{ij} (\hat{\sigma}^+_i\hat{\sigma}^-_j + \hat{\sigma}^-_i \hat{\sigma}^+_j),
\end{align}
with $J_{ij} \simeq J_0 / |i-j|^\alpha$. The power-law exponent $\alpha$ is controlled by adjusting the laser frequency of a three-tone laser field that manipulates electronic and transverse motional degrees of freedom of the ion chain \cite{Jurcevic2014}. For the current study, the exponent is tuned to $\alpha \approx 0.82$. 

In the numerical simulations, we used a spin-spin coupling matrix constructed from the experimental measurements to compare the experiment and theory. In the previous studies \cite{joshi2022observing, Kranzl2022}, the influence of the third tone, which is used for compensating the light-induced shifts on the spectator electronic levels, on the spin-spin coupling for long ion strings was neglected. Here, we revise the calculation of the $J_{ij}$ matrix and compare the experimentally measured spin-spin coupling matrix with the calculated one. For this calculation, we use the following expression after taking all motional modes ($2N$) and their respective coupling to each frequency component of the laser beam into account. Here, the coupling between the $i^{\text{th}}$ and $j^{\text{th}}$ ion is described by 
\begin{equation}
    J_{ij}\! =\! \frac{\hbar k^2}{4 m} \sum_{n=1}^{2N} M_{i}^{(n)} M_{j}^{(n)}\! \left[ \frac{\Omega_i^{\text{(B/R)}} \Omega_j^\text{(B/R)} }{\omega_{\text{B/R}}^2-\omega_n^2}+\frac{1}{2}\frac{ \Omega_i^{{(\text{C})}} \Omega_j^{(\text{C})} }{\omega_{\text{C}}^2-\omega_n^2} \right],
\end{equation}
where $M_{i}^{(n)}$ is the normalised mode amplitude of the $i^{\text{th}}$ ion and the $n^{\text{th}}$ motional mode with mode frequency $\omega_n$ \cite{Porras2004}. $\omega_{\text{R/B/C}}$ and $\Omega_{\text{R/B/C}}$ are the detunings and Rabi frequencies of the laser beam from the two-level atomic transition. Here, subscripts R, B, and C denote three frequencies of the laser beam that contains red-detuned, blue-detuned, and compensation beams. $k$ is the wavenumber of the 729~nm laser beam and $m$ is the mass of the calcium ion. 

\begin{figure}[t]
	\centering
	\includegraphics[width=1\columnwidth]{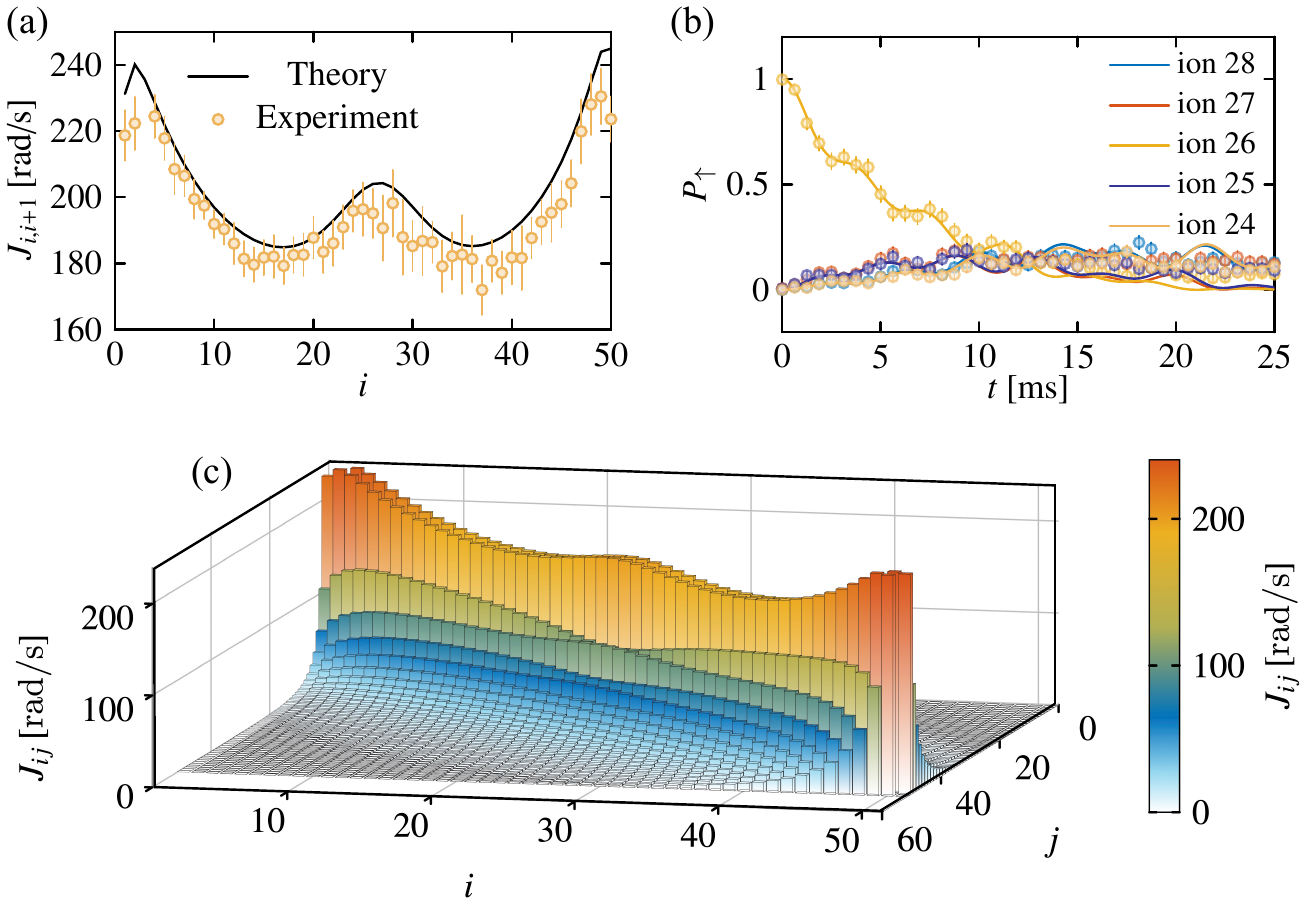}
	\caption{
 \textit{Effective interactions and spin dynamics for a 51-ion chain} (a) Experimentally measured nearest-neighbor interaction terms $J_{i, i+1}$, compared to theoretical calculations (solid line). (b) The quench dynamics of a single spin initialized to a spin-up state in the middle of the ion chain, while other spins initialized to a spin-down state, under the engineered flip-flop type interaction plotted in discs and solid lines are numerical results. (c) Theoretically calculated interaction matrix for the experimental parameters.}
 \label{fig:JijMatrix}
 \end{figure}

In our experiments, the laser is detuned by $\pm(\omega_{\text{COM}}+2\pi\times25~\text{kHz})$ from the carrier transition, while the center-of-mass mode frequency is $\omega_{\text{COM}} = 2\pi\times2.93$~MHz. Experimentally measured nearest-neighbor terms of the engineered spin-spin interaction are shown in Fig.~\ref{fig:JijMatrix}~(a), where solid lines are theory results. The engineered spin-spin interaction is then used to experimentally examine the quench dynamics of a single spin initialized to spin up in the middle of the ion chain while having all others in the spin-down state. The flip-flop interaction coherently drives the excitation to other locations of the ion chain while keeping the total magnetization conserved. The experimental results are shown in Fig.~\ref{fig:JijMatrix}~(b), where solid lines are numerical results. The full numerically simulated $J_{ij}$ matrix for the experimentally measured parameters is shown in Fig.~\ref{fig:JijMatrix}~(c).  

\section{System characterization} 
\label{se:StatePrep}

In our experimental platform, the input state is prepared with a global $X(\pi/2)$ gate over all ions, followed by an operation with a far-detuned laser beam that is tightly addressed over even ion sites out of all 51 ions while performing light shift gates. The unaddressed ions are prepared to spin down, while addressed ions are prepared to spin up after applying another global $X(-\pi/2)$ gate. The state preparation is subjected to inevitable dephasing processes, thus a spin-echo sequence is employed while splitting the sequential addressing operation into two parts to improve the state preparation. The state preparation fidelity $\mathcal{F}=|\braket{\downarrow \uparrow \downarrow \uparrow...|\psi_0}|^2$ is measured to be 0.75(7) for the whole ion chain, which corresponds to a single particle state preparation fidelity of $0.994(2)$. In the fidelity estimation, the error bars account for fluctuations over different days of measurement outcomes. An important remark: here, we perform direct fluorescence measurements (i.e. the measurements on the $z$ basis) and detect the individual ion on the EMCCD and compare the measured bit-string to the ideal state to calculate the fidelity. From our independent measurements, the detection error is measured to be smaller than $10^{-3}$ per particle, thus the drop in the present fidelity can be assigned to the state preparation. 

In addition to the state preparation error discussed above, there are also measurement errors in our experimental system: here, we discuss the results of two-qubit tomography performed for all nearest-neighbor pairs of the N{\'e}el state (the input state for the variational optimization). The average fidelity of the reconstructed two-qubit state is estimated to be $\mathcal{F_\text{two-qubit}}=0.980(5)$, i.e. $\mathcal{F_\text{single-qubit}}=0.989(5)$. This estimated fidelity also accounts for the state preparation discussed previously. To examine the major source of error in the measurements, we perform direct single-qubit tomography of the same input state while utilizing only global $x$, $y$, and $z$ base measurements, in contrast to the two-qubit case where local rotations are performed to account for all 9 bases measurements. Here, the average single-qubit state fidelity is estimated to be $0.994(3)$, which is higher than the estimates from the two-qubit tomography reconstruction. In summary, these analyses imply that in our system the leading errors, while carrying out measurements into different bases, arise from the local rotations which are performed with the help of a tightly focused beam.

\section{Entanglement Hamiltonian tomography and error mitigation} \label{sec:eht}
To reconstruct a reduced density matrix $\hat{\rho}_A$ from experimental data, and to gain insight into its entanglement structure, we use the procedure of entanglement Hamiltonian tomography (EHT) as introduced in Ref.~\cite{EHT}. Here, the reduced density matrix is assumed to be of the form 
\begin{align}\label{eq:expHA}
\rho_A(\betavec) = \exp \left( -\sum_{j\in A} \beta_j \hat{h}_j \right)/Z_A(\betavec).
\end{align}
The parameters $\beta_j$ are free variables to be fitted to the data, and $Z(\betavec) = \text{Tr}[\exp ( -\sum_{j\in A} \beta_j \hat{h}_j) ]$ is a constant that ensures trace normalization. The operators
\begin{align}
\hat{h}_j = \frac{J}{2} (\hat{S}^+_{j} \hat{S}^-_{j+1} + \mathrm{H.c.}) + \Delta \hat{S}^z_{j} \hat{S}^z_{j+1}
\end{align}
are associated with a link between two adjacent sites $j, j+1 \in A$, and form a decomposition of the Heisenberg Hamiltonian, i.e., such that $\hat{H} = \sum_j \hat{h}_j$.

State preparation and measurement errors in the experiment can be accounted for by applying a quantum operation consisting of a depolarizing map with rate $p_1$, and spontaneous emission from $\ket{\uparrow}$ to $\ket{\downarrow}$ with rate $p_2$, of the form
\begin{align}
\mathcal{D}_p[\rho_A(\betavec)] = \prod_{i=1}^{N_A} \mathcal{D}_p^{(i)} \rho_A(\betavec),
\end{align}
where $\mathcal{D}_p^{(i)}$ is a quantum operation, applied to particle $i$ of a density matrix $\hat{\rho}$:
\begin{align}\label{eq:Di}
\mathcal{D}_p^{(i)} \hat{\rho} = \sum_k E_k \hat{\rho} E_k^\dagger,
\end{align}
with the following $E_k$ acting on particle $i$:
\begin{align}
E_0 &= \sqrt{1 - 3p_1 / 4} \ket{\downarrow}\bra{\downarrow} + \sqrt{1 - 3p_1 / 4 - p_2} \ket{\uparrow}\bra{\uparrow}  \\
E_1 &= \sqrt{\frac{p_1}{4}} \hat{\sigma}^x, \ \ 
E_2 = \sqrt{\frac{p_1}{4}} \hat{\sigma}^y, \ \ 
E_3 = \sqrt{\frac{p_1}{4}} \hat{\sigma}^z,\\
E_4 &= \sqrt{p_2} \hat{\sigma}^- = \sqrt{p_2} \ket{\downarrow}\bra{\uparrow}
. \label{eq:Di_end}
\end{align}
The rates $p_1$ and $p_2$ are calibrated through an analysis of the total magnetization of the system after applying the state preparation circuit. Since the circuit conserves the magnetization, any deviations from the intended initial state magnetization can be attributed to the decoherence channel, and the values of $p_1$ and $p_2$ can be determined from the variance and mean of the magnetization.

Experimental data is collected as bit strings measured in the computational (Z) basis after applying unitary basis rotations 
\begin{align}\label{eq:U}
U^{(\boldsymbol{\alpha})} = \bigotimes_{j \in A} u_j^{(\alpha_j)}
\end{align}
where $\hat{u}_j^{(\alpha_j)}$ are single particle basis rotations applied at site $j$, rotating from the $\alpha_j = x, y, z$ basis to the $Z$ basis. We use $3^5$ different measurement settings $\bm{\alpha}$, corresponding to a tomographically complete basis set for all contiguous 5-site subsystems of the 51 ion chain (and in particular of subsystem A). While these measurements are not tomographically complete for larger subsystems, they still yield enough information to reliably fit the restricted ansatz Eq.~(\ref{eq:expHA}). In each basis, we take $10^2$ measurements, each measurement comprising a single bit string in the computational basis.

The free parameters $\beta_\li$ of the ansatz (\ref{eq:expHA}) are subsequently fitted to the acquired data in a least-squares sense, i.e., minimizing a cost function
\begin{align} \label{eq:cost}
\chi^2 \!= \!\sum_{\alphavec} \sum_{\boldsymbol{s}} \left[ P^{(\boldsymbol{\alpha})}_{\boldsymbol{s}} \!-\! \tr\left( \mathcal{D}_p [\rho_A(\betavec)] U^{(\alphavec)}\ket{\boldsymbol{s}}\bra{\boldsymbol{s}}U^{(\alphavec)\dagger} \right) \right] ^2, 
\end{align}
where $P^{(\boldsymbol{\alpha})}_{\boldsymbol{s}}$ is the experimentally observed probability for measuring bit string $\boldsymbol{s}$ after applying the basis transformation $\alphavec$.
This cost function represents the difference between the observed frequencies of occurrence of the bit strings, and the corresponding expectation values from the density matrix ansatz $\mathcal{D}_p [\rho_A(\betavec)]$ for a particular choice of parameters $\betavec$.

Finally, we note that the EHT procedure effectively provides a built-in method for error mitigation, i.e., it is possible to filter out the effects of decoherence. Namely, since the contributions of decoherence are explicitly fitted in the form of the quantum operation $\mathcal{D}_p$, we are thus able to isolate the coherent part $\rho_A({\betavec})$ from the experimental data. 

\section{Data post-processing} \label{methods:pstproc}
The XXZ-model studied in the main text exhibits a global $\mathbb{Z}_2$-symmetry
\begin{align}
[\hat{H}, \hat{\mathcal{P}}] = 0 \  \text{with} \ \hat{\mathcal{P}} = \bigotimes_i \hat{S}_i^x
\end{align}
resulting in the situation that eigenstates with opposite total magnetization $\ket{\Phi_n}$ and $\hat{\mathcal{P}}\ket{\Phi_n}$ are degenerate. The variational circuits used in the experiment (see Appendix~\ref{sec:vqs}) conserve total magnetization, hence starting with an initial state $\ket{\Psi_0}$ of given magnetization, the circuit is only able to prepare approximations to one of the states $\ket{\Psi_G}$ or $\hat{\mathcal{P}}\ket{\Psi_G}$.  

The original prediction of BW, however, is only valid for systems with a \emph{unique} ground state. The ground state (GS) degeneracy of the XXZ model, on the other hand, depends on the total number of sites $N$. For even $N$ the GS is unique with magnetization $M=0$, while for odd $N$ the GS is two-fold degenerate with $M=\pm 1$. In the limit $N\rightarrow \infty$, this distinction vanishes in the sense that the corresponding reduced density matrices $\rho_A$ converge to a single result. In order to test the BW prediction for the experimental system with an \emph{odd} number of sites, we approximate a pure ground state with mean magnetization given by the superposition $|\Psi_G \rangle +\hat{P} |\Psi_G \rangle$, where $|\Psi_G \rangle$ is one of the $M=\pm 1$ ground states. Explicitly, we calculate observables instead for the mixture $\ket{\Psi_G}\bra{\Psi_G} + \hat{\mathcal{P}}\ket{\Psi_G} \bra{\Psi_G}\hat{\mathcal{P}}$. Independent of the experimental analysis, we have numerically confirmed that this procedure converges to the correct expectation values in the bulk in the limit $N \rightarrow\infty$.

\section{Verification}
\label{Methods:XP}

To verify the learning procedure, we determine a (mixed-state) fidelity between the experimental quantum state under study, described by the density matrix $\rho_{\rm exp} \equiv \rho_1$, and the reconstructed density matrix from EHT,  $\rho_A(\betavec) \equiv \rho_2$. In particular, we analyze the reconstructed density matrix in terms of two different Hilbert-Schmidt fidelities defined in \cite{Liang2019}, given by the \emph{maximum} fidelity
\begin{align}
\mathcal{F}_{\textrm{max}}(\rho_{1},\rho_{2})=\frac{\tr(\rho_{1}\rho_{2})}{\max\{\tr(\rho_{1}^{2}),\tr(\rho_{2}^{2})\}},
\label{eq:Fmax}
\end{align}
and the \emph{geometric mean} fidelity
\begin{align}
\mathcal{F}_{\textrm{mean}}(\rho_{1},\rho_{2})=\frac{\tr(\rho_{1}\rho_{2})}{\sqrt{\tr(\rho_{1}^{2}) \tr(\rho_{2}^{2})}},
\label{eq:Fmean}
\end{align}
which both measure the overlap between $\rho_{1}$ and $\rho_{2}$, normalized by their purities. 

As shown in Ref.~\cite{ElbenXPlatform}, terms of the form $\tr(\rho_i\rho_j)$ for $i,j=1,2$, as occurring in Eqs.~\eqref{eq:Fmax}-\eqref{eq:Fmean}, can be evaluated from outcomes of measurements performed in sufficiently many measurement bases. Whereas Ref.~\cite{ElbenXPlatform} suggests the use of randomized measurement bases~\cite{elben2023randomized}, we use here the set of tomographically complete Pauli measurements for all contiguous subsystems of size 5, i.e. the same measurement bases used for the EHT protocol described above.

Specifically, we denote by $P^{(1)}_\alphavec(\boldsymbol{s})$ the frequency of having observed a particular bitstring $\boldsymbol{s}$ in the experiment (where $\rho_{\rm exp}$ is realized and Pauli basis rotation $\hat{U}^{(\alpha)}$ has been applied) and $P^{(2)}_\alphavec(\boldsymbol{s})= \tr\left( \rho_A(\betavec) U^{(\alpha)}\ket{\boldsymbol{s}}\bra{\boldsymbol{s}}U^{(\alpha)\dagger} \right)$, i.e. the expectation value of observing bitstring $\boldsymbol{s}$ in the reconstructed state $\rho_A(\betavec)$. Then, we obtain  the overlap $\tr(\rho_i\rho_j)$  for $i=1,j=2$ and purities $\tr(\rho_i\rho_j)$ for $i=j=1,2$ via~\cite{ElbenXPlatform}
\begin{align}
\tr(\rho_i\rho_j)= \label{eq:ovl} 
\frac{2^{N_{A}}}{N_\alphavec}\sum_\alphavec \sum_{\boldsymbol{s},\boldsymbol{s}'}(-2)^{-\mathcal{D}[\boldsymbol{s},\boldsymbol{s}']}{P_{\alphavec}^{(i)}(\boldsymbol{s})P_{\alphavec}^{(j)}(\boldsymbol{s}')},
\end{align}
where the Hamming distance $\mathcal{D}[\boldsymbol{s},\boldsymbol{s}']$ between two strings $\boldsymbol{s}$ and $\boldsymbol{s}'$ is defined as the number of entries where $s_{k}\neq{s}'_{k}$, i.e.\ $\mathcal{D}[\boldsymbol{s},{\boldsymbol{s}}']\equiv\#\left\{ k\in \{1,\dots, N_A\}\,|\,s_{k}\neq{s}'_{k}\right\} $  (see also Refs.~\cite{FlammiaDirectFidelity, PoulinDirectFidelity}). 

Eq.~(\ref{eq:ovl}) provides a direct experimental verification of the reconstructed density matrix via the Hilbert-Schmidt fidelity, requiring no further theory input such as simulations, and can be evaluated from the same type of measurements employed for EHT. Importantly, however, the measurements used for fidelity estimation should be independent from those used in EHT, to avoid false correlations and biasing. We, therefore, split the total dataset for each quantum state into two, where one-half of the data is used to reconstruct $\rho_A(\betavec)$, and the other half is subsequently used to evaluate the fidelity (\ref{eq:Fmax}).
To evaluate the purity $\tr(\rho_1\rho_1)$, the verification data set is split once more into two sets to perform an unbiased estimation. That is, in Eq.~(\ref{eq:ovl}), we evaluate $P_{\alphavec}^{(j)}(\boldsymbol{s})$ from one dataset, and $P_{\alphavec}^{(j)}(\boldsymbol{s}')$ from the other.

Finally, since the measurement basis set is only tomographically complete for contiguous subsystems of size 5, and since we have taken only a limited number of measurements, we have found that the fidelity estimation becomes inaccurate for subsystems larger than 5. For subsystems of size $L_A > 5$ we therefore compute the averaged 5-site fidelity for all contiguous sub-subsystems contained in the subsystem, as a measure of the fidelity of the total subsystem.

\section{Variational state preparation} 
\label{sec:vqs}

The experiment prepares variational quantum states~\cite{Cerezo2021} by alternatingly applying two types of unitaries. The first type of operation consists of an entangling operation of the form
$
\hat{U}_{XY}(\theta) = \exp(-\mathrm{i} \theta \hat{H}_{XY} )$,
which applies for a variable duration $\theta$ the native interaction Hamiltonian described in Eq. \eqref{eq:xy}. The second type of operation consists of single particle rotations, applied to each second site, of the form
$
\hat{U}_Z(\theta) = \exp\left(-\mathrm{i} \sum_{k = 1}^{\lfloor N/2 \rfloor} \hat{\sigma}^z_{2k} \theta/2\right)$. 
Starting from an initial N\`eel state $\ket{\psi_0} = \ket{\downarrow \uparrow \downarrow  \ldots}$, the variational quantum states are thus of the form
\begin{align}
\ket{\Psi(\boldsymbol{\theta})} = \hat{U}_Z(\theta_M) \hat{U}_{XY}(\theta_{M-1}) \cdots \hat{U}_Z(\theta_2) \hat{U}_{XY}(\theta_1) \ket{\psi_0}.
\end{align}

The parameters $\boldsymbol{\theta}$ of the variational state are optimized in a feedback loop with a classical computer running an optimization algorithm that attempts to minimize the expectation value of the energy of the state  under the Heisenberg Hamiltonian $\hat{H}$. From 30 measurements in each of the $X, Y, Z$ bases, the quantity $\bra{\Psi(\boldsymbol{\theta})} \hat{H} \ket{\Psi(\boldsymbol{\theta})}$ is estimated, serving as a cost function for the classical optimizer. We use a variant of the SPSA algorithm \cite{Kandala2017}, enhanced with a Gaussian process surrogate model \cite{RasmussenWilliams}. The search is warm-started with an initial guess consisting of optimal parameters from a numerical exact optimization for 13 particles. The variational optimization on the experiment serves to refine these parameters. We reach energies, normalized to the total spectral range, of $0.048(2)$ for $\Delta = 1$ and $0.056(8)$ for $\Delta = 1.7$. 

For the `heated states' discussed in the main text, we apply one additional quench with the native entangling operation $\hat{H}_{XY}$. This quench is applied to the initial state, before running the variational circuit with optimal parameters. We found that quenches to the initial state are more effective at heating the state than similar length quenches applied \textit{after} executing the circuit. The duration of the heating quenches is $1.87$~ms and $1.5$~ms for $\Delta = 1$ and for $\Delta = 1.7$, respectively.
For these quenches, the measured energies on the quantum simulator are $0.37(1)$ and $0.197(4)$.

We note that there also exist deterministic state preparation protocols for eigenstates of the XXZ chain, making use of the integrability of the model, in terms of so-called algebraic Bethe circuits~\cite{sopena2022algebraic}. Given the requirement of a universal gate set to implement these circuits, we have employed simpler short-depth variational circuits here.

\section{Bisognano-Wichmann theorem and extensions} \label{sec:BWmethods}

\subsection{The Entanglement Hamiltonian in Quantum Field Theories}
Entanglement properties of quantum many-body systems can rarely be described analytically. A notable exception is given by the Bisognano-Wichmann (BW) theorem~\cite{BW1, BW2, witten2018aps}, which applies to the ground state $|\Omega\rangle$ of any relativistic quantum field theory (RQFT) in $(d+1)$-dimensional Minkowski space. Given the underlying Hamiltonian $\hat{H} = \int {\sf d}^{d}x \, \mathscr{H}(x)$ where $\mathscr{H}(x) = T^{00}(x)$ is determined by the energy-momentum tensor $T^{\mu \nu}(x)$, the reduced density operator as given in Eq.~\eqref{eq:BWQFT} of the main text
	can be calculated exactly for the special case of a bi-partition of space into two halves, $\mathbb{R}^{d} = A \cup B$ with $A = \{x = (x_1, \dots, x_{d}) \in \mathbb{R}^{d} | x_1>0\}$. The corresponding EH takes the form of a `deformation' of $\hat{H}$,
	\begin{align}
		\tilde{H}_A =  \int_{x\in A} {\sf d}^{d}x \, \beta(x) \mathscr{H}(x) - F\;, 
	\end{align}
	with a linearly increasing `local inverse temperature'
	\begin{align}\label{eq:BW_QFT}
		\beta(x) = \beta^\text{BW}(x) = 2\pi x_1, 
	\end{align}
	and a normalization constant $F$.
	
	The classic result of BW can be extended when the QFT exhibits conformal invariance~\cite{casini2011towards,cardy2016entanglement}, which allows to map the half-space into other regions $A \subset \mathbb{R}^{d}$. In particular, for a solid sphere of radius $R$, i.e. $A =  \{x \in \mathbb{R}^{d} | x^2  \le R^2\}$, the EH is again local with
	\begin{align}\label{eq:BW_CFT}
		\beta(x) = \beta^{\text{CFT}}(x) = 2\pi \frac{(R^2 - x^2)}{2R} \;.
	\end{align}
	We emphasize that the predictions in Eqs.~\eqref{eq:BW_QFT} and \eqref{eq:BW_CFT} are valid in arbitrary spatial dimensions $d$ and only require Lorentz or conformal invariance of the QFT, respectively.
	
	Translating these analytical results from the continuum to a spatial lattice suggests approximate linear and parabolic deformations such as
	\begin{align}
		  \beta^\text{BW}(x)\quad \rightarrow &\quad \beta^{\text{BW}}_n \sim n\;, && n=0,1,2, \dots \;, \\
		  \beta^\text{BW}(x)\quad \rightarrow &\quad \beta^{\text{CFT}}_n \sim n(N-n)\;, && n=-N, \dots N
	\end{align}
	written here for ${d}=1$ for simplicity [see also Eq.~\eqref{eq:deformation}]. 
	
	For disconnected intervals as studied in the main text, no generally applicable extension of the BW theorem is presently known. For reference, we state here the result for a massless Dirac field on a line with $A= A_+ \cup A_- \subset \mathbb{R}$ and two intervals $A_\pm = (\pm a,\pm b)$~\cite{casini2009reduced,eisler2022local}
	\begin{align}
	    \tilde{H}_A = \int_{x\in A} dx \left[ \beta_\text{loc.}(x) \mathscr{H}(x) +\beta_\text{bi-loc.}(x) \mathscr{H}_\text{bi-loc.}(x,x_c(x)) \right]\;.
	\end{align}
	Here, $\mathscr{H}_\text{bi-loc.}(x,x_c)$ is a bi-local operator that connects every $x \in A_\pm$ with a conjugate point $x_c(x)= -ab/x \in A_\mp$, and the spatial deformations take the form
	\begin{align}
	    \beta_\text{loc.}(x) &= \frac{(b^2 - x^2)(x^2 - a^2)}{ 2(b- a)(ab+x^2)} \;, \\
	    \beta_\text{bi-loc.}(x) &=  \frac{a b}{x(ab+x^2)} \beta_\text{loc.}(x)  \;.
	\end{align}
	Note that the local function $\beta_\text{loc.}(x)$ interpolates between the expected parabolic shapes for a single subsystem when $A_+$ and $A_-$ touch ($b=R$ with $a\rightarrow 0$) and two independent subsystems when $A_+$ and $A_-$ are far away ($b = a+ 2R$ with $a\rightarrow \infty$), while the bi-local function $\beta_\text{bi-loc.}(x)$ vanishes in both limits.

\subsection{Applicability to general low-energy states} \label{sm:200}
\begin{figure*}[t]
	\centering
	\includegraphics[width=0.75\textwidth]{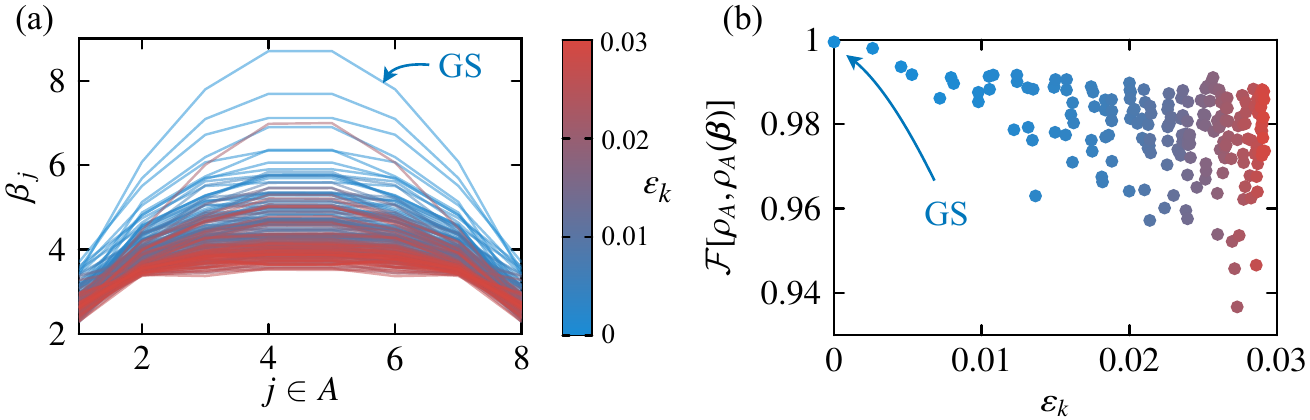}
	\caption{ \textit{Numerical analysis of the Entanglement Hamiltonian of the lowest 200 eigenstates of a 51-site XXZ chain} (a) Local temperature profiles of the EH for subsystems of $L_A=9$ sites in the center of the spin chain for the individual eigenstates. The inverse temperature profiles are colored according to their eigenenergies, which are measured in units of the total spectral range. (b) Verification of the local structure of the EH via computing Uhlmann fidelities between the density matrices $\rho_A(\boldsymbol{\beta})$, reconstructed from EHT, and the exact density matrices $\rho_A$ obtained via performing a Schmidt decomposition on the individual eigenstates $\ket{\Phi_k}$ with $k = 1 \dots 200$. GS denotes the temperature profile as well as the fidelity for the ground state wave function.
	}
	\label{supfig:200}
\end{figure*}

Predictions about the spatial structure of the Entanglement Hamiltonian (EH) from RQFT and Conformal Field Theory (CFT) apply to \emph{ground states} of local theories (see main text). In the present paper, we study the EH for subsystems of variationally prepared quantum many-body states on a trapped-ion quantum simulator. We find that these states can be represented as superpositions of a finite number of low-lying eigenstates of the target model. While numerical verification of the local structure of the EH has been established for ground states of quantum lattice models \cite{dalmonte2022entanglement}, it is less clear if such findings hold true for low-lying excited states or their superpositions. In the following, we study the EH's spatial profile in low-lying eigenstates of a 51-site XXZ chain and verify its local structure using fidelity estimations with respect to the exact density matrices.

Specifically, we compute excited states of the 51-site XXZ model using the Density Matrix Renormalization Group (DMRG) with matrix product states (MPS). In particular, for calculating the $k^\text{th}$ excited state, we modify the XXZ Hamiltonian $\hat{H}$ by adding projectors on the $k-1$ pre-computed eigenstates $\ket{\phi_k}$, with a constant weight factor $w$:
\begin{align}
\hat{H} \rightarrow \hat{H} + w \sum_{q<k} \ket{\phi_q} \bra{\phi_q}.
\end{align}
Within this scheme, DMRG minimizes the energy of $\hat{H}$ while simultaneously minimizing the overlap with all previously computed excited states. Using this strategy, we compute MPS representations for the lowest 200 excited states of the XXZ model. 

In Fig.~\ref{supfig:200}  (a) we performed EHT for subsystems of $L_A = 9$ sites for the lowest 200 excited states. As can be seen, each of the individual eigenstates exhibits a parabolically deformed EH, where the temperature profile for higher-lying states acquires a flat plateau in the bulk of the subsystem. 

In order to verify the validity of the local temperature profiles, in Fig.~\ref{supfig:200}  (b) we compute the Uhlmann fidelity $\mathcal{F}(\rho_1, \rho_2) = \left( \text{Tr} \sqrt{\sqrt{\rho_1} \rho_2 \sqrt{\rho_1}}\right)^2$ between the density matrices $\rho_A(\boldsymbol{\beta})$, obtained from EHT, and the exact density matrices $\rho_A = \text{Tr}_{\bar{A}} \left( \ket{\Phi_k} \bra{\Phi_k} \right)$. While the overall fidelities consistently exceed $\sim\,94\%$ for the lowest 200 eigenstates, Fig.~\ref{supfig:200}  (b) reveals that the spread in fidelity increases for higher-lying excited states. However, even at the highest energies investigated, eigenstates can be found where the reduced density matrix can be described by a local EH with up to $\sim$ 99\% fidelity.

\section{Numerical justification of the data post-processing procedure}

\begin{figure}[t]
	\centering
	\includegraphics[width=0.45\textwidth]{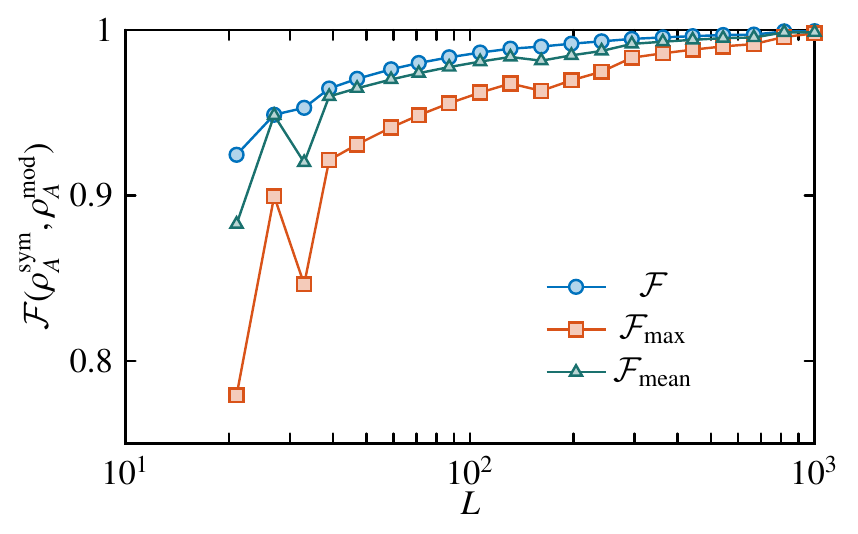}
	\caption{
 Uhlmann fidelity $\mathcal{F}$, maximum fidelity $\mathcal{F}_\text{max}$ and geometric-mean fidelity $\mathcal{F}_\text{mean}$ (see Appendix \ref{Methods:XP}) of the `correct' density matrix $\rho_A^\text{sym}$ with respect to the modified density matrix $\rho_A^\text{mod}$ as a function of total system size $L$, for a 5-site subsystem from the center of the spin chain. The plot illustrates that contributions to the density matrix $\rho_A^\text{sym}$ originating from off-diagonal coherences in the state $\ket{\Psi^\text{sym}}$ vanish in the limit $L\rightarrow \infty$. 
	}
	\label{supfig:justi}
\end{figure}

As discussed in the main text, in our experiment we prepare approximate ground states $\ket{\Psi_G}$ of the XXZ model for $N=51$ sites in different parameter regimes. These states have good quantum numbers with respect to total magnetization $\hat{M} = \sum_j \hat{S}_j^z$ but are not eigenstates of the global $\mathbb{Z}_2$ symmetry operator $\hat{\mathcal{P}} = \bigotimes_j \hat{S}_j^x$. This is because our variational circuits prepare states with a fixed finite magnetization $M\ne 0$, while the global $\mathbb{Z}_2$ operation  changes the magnetization $M$ to $-M$. In Appendix \ref{sec:BWmethods}, we discuss a data post-processing technique that effectively simulates the presence of a state of the form $\ket{\Psi_G^\text{sym}} = \frac{1}{\sqrt{2}} \left(\ket{\Psi_G} + \hat{\mathcal{P}}\ket{\Psi_G} \right)$ in experimental measurements, restoring the global $\mathbb{Z}_2$ symmetry and facilitating the investigation of the BW predictions through an analysis of a unique ground state (see Methods). This is achieved via an approximation, modifying the experimentally obtained sample data to make them appear as if they originate from a state of the form
\begin{align}\label{eq:rhomod}
\rho_A^\text{mod} = \frac{1}{2} \left( \rho_A + \hat{\mathcal{P}}_A \rho_A \hat{\mathcal{P}}_A \right),
\end{align}
with $\hat{\mathcal{P}}_A = \bigotimes_{j \in A} \hat{S}_j^x$ and $\rho_A$ the reduced density matrix of a subsystem of $\ket{\Psi_G}$. In the following, we justify this procedure via numerical simulations and show that it converges to the `correct' density matrix
\begin{align}
\rho_A^\text{sym} = \text{Tr}_{\bar{A}} \left( \ket{\Psi_G^\text{sym}} \bra{\Psi_G^\text{sym}} \right),
\end{align}
in the limit $L \rightarrow \infty$, where $L$ denotes the total number of particles.

To verify the procedure, we compute $\rho_A^\text{sym}$ of a bulk region of the superposition state $\ket{\Psi_G^\text{sym}}$ and compare it to the modified density matrix $\rho_A^\text{mod}$ of Eq.~(\ref{eq:rhomod}). Expanding the reduced density matrices of a subsystem $A$ in a basis of Pauli strings $\hat{\sigma}_j^{\alpha_j}\hat{\sigma}_{j+1}^{\alpha_{j+1}}\dots \hat{\sigma}_{j+L_A-1}^{\alpha_{j+L_A-1}}$, we can find an analytical expression for the difference between $\rho_A^\text{sym}$ and $\rho_A^\text{mod}$, given by
\begin{align} \label{eq:rhodiff}
\begin{split}
\rho_A^\text{sym} - \rho_A^\text{mod} &=  \sum_{\{ \alpha_j \}} \bra{\Psi_G} \{ \hat{\sigma}_j^{\alpha_j}\hat{\sigma}_{j+1}^{\alpha_{j+1}}\dots \hat{\sigma}_{j+L_A-1}^{\alpha_{j+L_A-1}}, \hat{\mathcal{P}} \} \ket{\Psi_G} \\ &\times \hat{\sigma}_j^{\alpha_j}\hat{\sigma}_{j+1}^{\alpha_{j+1}}\dots \hat{\sigma}_{j+L_A-1}^{\alpha_{j+L_A-1}}
\end{split}
\end{align}
where $\{\cdot, \cdot \}$ denotes the anticommutator and $\hat{\sigma}_j^{\alpha_j} \in [\hat{\mathbb{1}}_j, \hat{\sigma}_j^x, \hat{\sigma}_j^y, \hat{\sigma}_j^z]$. The expectation values appearing in Eq.~(\ref{eq:rhodiff}) contain Pauli strings of the form $\hat{\sigma}_{1}^{x} \cdots \hat{\sigma}_{j-1}^{x} \hat{\sigma}_j^{\alpha_j}\hat{\sigma}_{j+1}^{\alpha_{j+1}}\dots \hat{\sigma}_{j+L_A-1}^{\alpha_{j+L_A-1}} \hat{\sigma}_{j+L_A}^{x} \cdots \hat{\sigma}_{L}^{x} $ which span the whole spin chain. 
Intuitively, expectation values of large Pauli strings vanish for $L \gg$, which we numerically verify in Fig.~\ref{supfig:justi}.

In particular, Fig.~\ref{supfig:justi}  illustrates different density matrix fidelities (see Appendix \ref{Methods:XP}) between $\rho_A^\text{sym}$ and $\rho_A^\text{mod}$ for density matrices computed for a subsystem of size $L_A = 5$ from the center of the spin chain as a function of the total system size $L$. As can be seen, $\rho_A^\text{sym}$ and $\rho_A^\text{mod}$ become approximately identical at $L\sim 10^3$, indicating the vanishing of expectation values of large Pauli stings as $L \rightarrow \infty$. Contributions to the density matrix $\rho_A^\text{sym}$ that stem from off-diagonal coherences in $\ket{\Psi^\text{sym}}$ thus vanish for $L\rightarrow \infty$. This allows us to replace the coherent superposition state $\ket{\Psi^\text{sym}}$ by an incoherent mixture $\ket{\Psi_G}\bra{\Psi_G} + \hat{\mathcal{P}}\ket{\Psi_G} \bra{\Psi_G}\hat{\mathcal{P}}$ for large system sizes, for which reduced density matrices take the form of Eq.~(\ref{eq:rhomod}).

\section{Spatial structure of entanglement eigenstates}
\begin{figure}[t]
	\centering
	\includegraphics[width=0.45\textwidth]{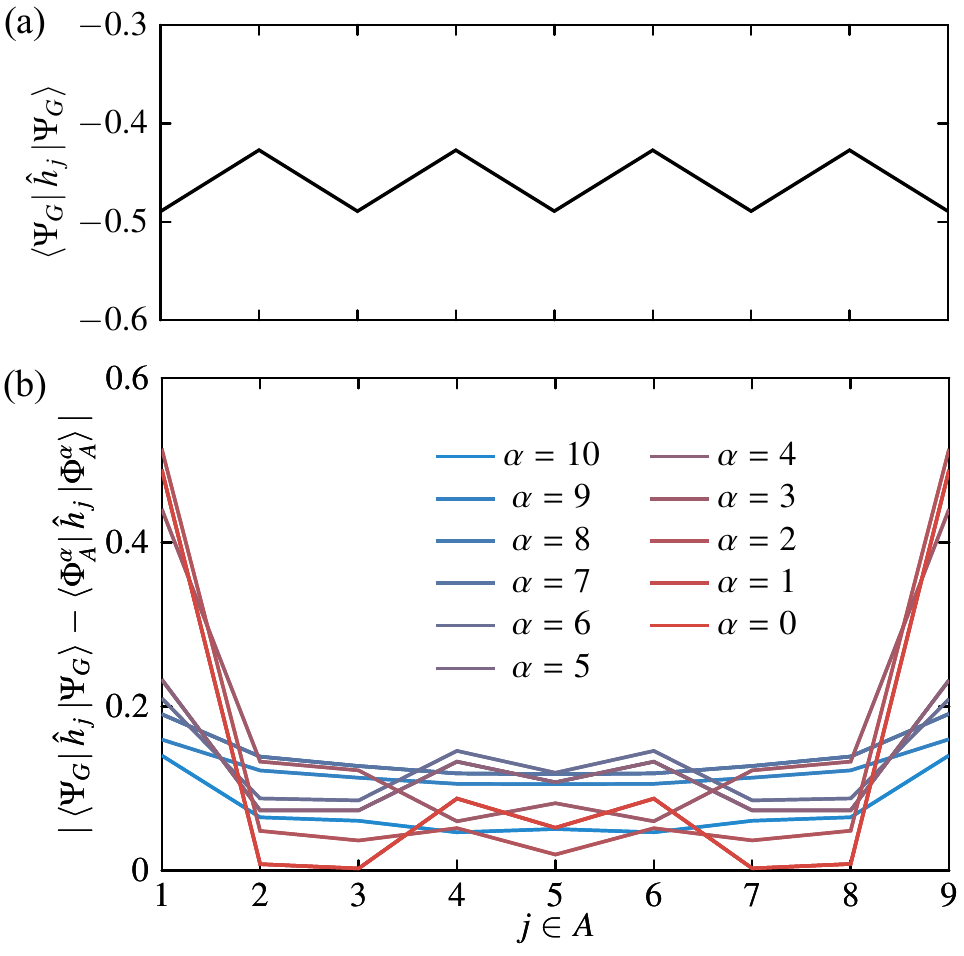}
	\caption{\textit{Energy density of entanglement eigenstates in comparison to the ground state.} (a) Energy density in the ground state of a 100-site XXZ model at the isotropic point $J=\Delta=1$. (b) Absolute difference between energy densities of the ground state wave function and entanglement eigenstates of a 10-site subsystem from the bulk of the chain. Note that the excess energy in the dominant Schmidt vectors $\ket{\Phi_A^{\alpha}}$ with $\alpha = 0, 1, 2$ with respect to the ground state $\ket{\Psi_G}$ is concentrated close to the entanglement cut.}
	\label{supfig:schmidt}
\end{figure}

In this section we provide further evidence for the interpretation of the deformation parameters $\beta_j$ in the EH as a ``local inverse (entanglement) temperature''. To be explicit, we consider again the EH
$\tilde{H}_A = \sum_{j\in A} \beta_j h_j$ with a finite subsystem $A$ in the bulk for the ground state $\ket{\Psi_G}$ of $\hat{H}_{\text{XXZ}} = \sum_j h_j$ as in the main text. The fact that $\rho_A \propto e^{-\tilde{H}_A}$ is a Gibbs state with $\beta_j$ small close to the entanglement cut and larger towards the middle of the subsystem intuitively suggests that the ``dominant contributions to the entanglement live close to the boundary''. 

To make this intuition more precise, we write 
\begin{align}
\rho_A = \frac{1}{Z_A} e^{-\tilde{H}_A} = \sum_\alpha e^{-\xi_\alpha} \ket{\Phi_A^{\alpha}} \bra{\Phi_A^{\alpha}}
\end{align}
with $\xi_\alpha$ and $\ket{\Phi_A^{\alpha}}$ the eigenvalues and -states of $\tilde{H}_A$, i.e. the entanglement spectrum and the Schmidt vectors, respectively. The dominant contribution to the entanglement is thus attributed to the Schmidt vectors with the smallest eigenvalues. We expect that these vectors carry excitations that are dominantly supported close to the entanglement cut.

We reveal this anticipated spatial structure by calculating the average energy density $\langle \hat{h}_j \rangle$ and $j\in A$ (see Fig.~\ref{supfig:schmidt}). In Fig.~\ref{supfig:schmidt}a, we show the energy density for $\ket{\Psi_G}$, which exhibits a homogeneous profile (up to a 2-site unit cell effect) as expected for a translationally invariant system. In contrast, the energy density in the low-lying Schmidt vectors is strongly inhomogeneous as demonstrated in Fig.~\ref{supfig:schmidt} (b), where we plot the absolute difference in energy density in $\ket{\Phi_A^{\alpha}}$ w.r.t. $\ket{\Psi_G}$. We interpret the fact that this difference is largest close the entanglement cut as localized excitations in the dominant Schmidt vectors which are supported in this boundary region.

This spatial ``localization'' of entanglement finds its most prominent manifestation in the context of topological order. As originally pointed out by Li and Haldane~\cite{li2008entanglement}, the low-lying entanglement spectrum for a subsystem of  topologically ordered state carries a fingerprint of the structure associated to the edge state CFT, which is again due to dominant Schmidt vectors contributing excitations that are mainly supported close to the entanglement cut. Our findings suggest that such interpretations also apply in a more general context.

\bibliographystyle{apsrev4-1}    

\end{document}